\documentclass[%
 reprint,
superscriptaddress,
 amsmath,amssymb,
 aps,
pra,
]{revtex4-1}

\usepackage{graphicx}
\usepackage{dcolumn}
\usepackage{bm}
\usepackage{braket}
\usepackage{color}

\begin{document}

\preprint{APS/123-QED}

\title{Optimal control of two qubits via a single cavity drive in circuit quantum electrodynamics}

\author{Joseph L. Allen}
 \email{joseph.allen@surrey.ac.uk}
 \affiliation{%
 Advanced Technology Institute and Department of Physics\\
 University Of Surrey, Guildford GU2 7XH, United Kingdom
}%
 
 \author{Robert Kosut}
\affiliation{
 SC Solutions, Inc., 1261 Oakmead Parkway, Sunnyvale, CA 94085
}%

\author{Jaewoo Joo}%
\affiliation{%
 Advanced Technology Institute and Department of Physics\\
 University Of Surrey, Guildford GU2 7XH, United Kingdom
}%
\affiliation{School of Computational Sciences, Korea Institute for Advanced Study, Seoul 02455, Republic of Korea}


\author{Peter Leek}
\affiliation{
Clarendon Laboratory, Department of Physics\\
University of Oxford, Oxford OX1 3PU, United Kingdom}

\author{Eran Ginossar}
\affiliation{%
 Advanced Technology Institute and Department of Physics\\
 University Of Surrey, Guildford GU2 7XH, United Kingdom
}%


\date{\today}

\begin{abstract}
Optimization of the fidelity of control operations is of critical importance in the pursuit of fault-tolerant quantum computation. We apply optimal control techniques to demonstrate that a single drive via the cavity in circuit quantum electrodynamics can implement a high-fidelity two-qubit all-microwave gate that directly entangles the qubits via the mutual qubit-cavity couplings. This is performed by driving at one of the qubits' frequencies which generates a conditional two-qubit gate, but will also generate other spurious interactions. These optimal control techniques are used to find pulse shapes that can perform this two-qubit gate with high fidelity, robust against errors in the system parameters. The simulations were all performed using experimentally relevant parameters and constraints.
\end{abstract}

\maketitle


\section{\label{sec:level1}Introduction}

Fault-tolerant architectures for quantum computation require individual gate operations to be performed with high fidelity \cite{Aharonov2008}. This requires strict control of all the parameters in the system and robustness of the performed function with respect to any noise or uncertainties in the system. In superconducting circuits, experimental two-qubit gate fidelities $\mathcal{F}>0.99$ have recently been demonstrated \cite{Sheldon2016, Martinis2014, Barends2014, Ghosh2013a}, at the threshold for fault tolerance with a surface code architecture \cite{Fowler2009, Fowler2012}, but well below that required for a gate-model machine \cite{Aharonov2008, Nielsen2010}. Although some of the infidelity is due to decoherence, there are also errors due to fluctuations and inaccuracies in the driving field and measured system parameters. 

The systems are also required to be scalable, so that more qubits can be added in order to realize a full quantum computer \cite{DiVincenzo2000}. A crucial challenge in scaling up is to maintain high coherence, which becomes more difficult as more controls are added. Therefore, it can be ideal to keep the circuit complexity to a minimum in order to reduce the potential avenues for decoherence. Then, the problem becomes one of how to control the system with fewer sources. Control in quantum mechanics is an open problem and equivalent to a non-linear optimization, which is difficult to parametrize and solve analytically, but is an important problem for quantum computing.

In this paper, we investigate an all-microwave gate in circuit quantum electrodynamics (cQED) using a single microwave drive applied to a single cavity containing two transmon \cite{Koch2007a} qubits. The system dynamics are controlled by selecting the frequency, phase, and amplitude of the microwave control drive \cite{Rigetti2010, Groot2010, Groot2012}. Using only a single microwave drive for the control means the circuit complexity is kept to a minimum while also minimizing the external sources of noise. We use the sequential convex programming (SCP) algorithm developed in \cite{Kosut2013} to find pulses that are capable of implementing an entangling operation between two qubits in one cavity to a high degree of fidelity via the cross-resonance gate \cite{Rigetti2010, Chow2011}. This gate was first implemented in circuit QED using microwave drives local to each qubit and utilizes the qubit-qubit coupling via the common mode of the cavity. Using a fixed system and modifying only the drive pulses, we show that it is possible to implement the desired unitary with fidelities $\mathcal{F} > 0.9875$ using a two-level approximation for the qubits, and $\mathcal{F}>0.9639$ for a full multi-level transmon model. We also show that these fidelities are robust to uncertainty in the system parameters and incorporation of experimentally realistic pulse filtering, while also imposing constraints on the drive power.

Advanced pulses for control in quantum systems have been investigated previously, such as dynamical decoupling schemes which use trains of pulses to cancel out environmental noise and get rid of any dephasing on the qubit \cite{Viola1998, Viola1999}. DRAG (derivative removal via adiabatic gate) was designed in order to remove any leakage to the non-computational levels of the qubit, here two pulses are applied to the system with the second being the derivative of the first \cite{Motzoi2009, Chow2010}. Other examples include spline-shaped pulses for the resonator-induced phase gate where the pulse shapes are used to remove unwanted effects during the gate such as photon loss and residual entanglement between the qubits and cavity after the gate implementation \cite{Cross2015}, and SWIPHT (sppeding up wave forms by inducing phases to harmful transitions) that purposely drives the nearest harmful transition so that it undergoes one cyclic revolution while maximizing the fidelity to the desired unitary \cite{Economou2015}. While these methods have proven to be effective, they have not included any potential sources of error in the system which will inevitably be present in a real experiment. More recently, optimal control methods were used to find the minimum time to perform quantum operations in superconducting systems, in particular the cross-resonance gate was investigated for this purpose using the common method of dedicated qubit control drives \cite{Kirchhoff2017}. Work was also performed in superconducting circuits for a single global cavity drive, searching for the best parameter regime to perform single- and two-qubit gates in the shortest possible time and maximizing the fidelity to the gates of interest for the best parameter regime \cite{Goerz2016}. In these cases again potential error sources were not included in the investigation.

Quantum optimal control theory is one method for designing pulse shapes to perform required interactions \cite{Gross2007,Brif2010}. In particular, it has been used to develop algorithms for numerically designing pulse shapes that maximize a chosen fidelity function \cite{Navin2005,DeFouquieres2011, Ferrie2014, Machnes2015, Schirmer2011, Kosut2013, Goerz2014, Wilhelm2007}. It has shown success in designing pulses for single-qubit gates, two-qubit gates, and readout \cite{Ginossar2010}. There have also been applications of quantum optimal control algorithms towards high fidelity gates that are robust to errors in the system \cite{Kosut2013, Dong2014a, Goerz2014}, but these have focused either on single-qubit gates, assuming that the control came via changing the system parameters, for example, via flux tuning, or by direct driving of the qubit of interest.

The rest of the paper is organized as follows: In Sec. \ref{sec:CR} we define the Hamiltonian for two qubits in a three-dimensional (3D) cavity driven by a single microwave drive, then perform a set of transformations relevant to the system in order to derive the two-qubit interaction. We see that upon making these transformations that there will be some unwanted interaction terms that are being driven resonantly that will be harmful to the desired entangling operation. In Sec. \ref{sec:OCT} we describe the quantum optimal control method, and in particular we review the sequential convex programming algorithm and how it can be used to perform robust quantum optimal control. In Sec. \ref{sec:Results} we present our results and discussion, showing that using SCP we can design a pulse shape that can perform the desired entangling operation with high fidelity for both a two-level system and a multi-level system even in the presence of resonant unwanted rotations, errors in the system parameters, and pulse filtering.

\section{\label{sec:CR}All-Microwave Gate Using a Single Cavity Drive}

The cross-resonance gate is an all-microwave gate that utilizes coupling between two qubits to generate a two-qubit operation \cite{Chow2011, Rigetti2010}. For all-microwave gates the system is set up so that the qubits are far detuned from one another, and also from the cavity (i.e., the system is in the dispersive regime). This ensures that the qubits and cavity do not interact until some external control is applied, and will extend the lifetimes of the qubits when they are not being operated on. In the case where there are microwave drives local to each qubit, a microwave resonant with the target qubit would be applied directly to the other control qubit via a direct microwave line. Due to the coupling between the two qubits via the common mode of the cavity this would activate a two-qubit operation which could be used to generate entanglement. This system ideally requires a local drive on each qubit, but it is interesting to consider a case in which only a single drive is available that globally addresses both qubits. This could be an advantage in large-scale systems, as it reduces the number of required controls. 

One such simple architecture is to have a single drive coupled to the cavity and use this to perform all the control; this is relevant when the system consists of two-qubits in a 3D cavity. This has been shown to be of interest due to the long coherence times enabled by the design \cite{Paik2011}. In the case of the cross-resonance gate this would mean that the microwave drive that is resonant with the target qubit will not only activate the two-qubit operation, but will also cause unwanted rotations of the target qubit.

In superconducting circuits, one of the qubits of interest, and the focus of this work, is the transmon \cite{Koch2007a}. This is a charge-based qubit that is designed in such a way as to be less susceptible to charge noise, thus increasing coherence times. However, the trade off is that the system is less anharmonic and is thus not a true two-level system but is in fact multi-level. The Hamiltonian for two transmons coupled to a resonator with a single-microwave control is given by:

\begin{equation}
\begin{split}
    H= \hspace{1mm}&\omega_{r}a^{\dagger}a+\sum_{i=1,2}\sum_{j_{i}}\omega_{j_{i}}\ket{j_{i}}\bra{j_{i}}+\sum_{i=1,2}g_{i}(a^{\dagger}c_{i}+ac_{i}^{\dagger})\\
    &+[\varepsilon(t)a^{\dagger}e^{-i\omega_{d}t}+\varepsilon^{*}(t)ae^{+i\omega_{d}t}],
\end{split}
\end{equation}
\\
where $\omega_{r}$ is the cavity resonance frequency, $i$ denotes transmon $1$ and $2$, $\omega_{j_{i}}$ is the frequency of the $j^{\mathrm{th}}$ level of the $i^{\mathrm{th}}$ transmon $\ket{j_{i}}$, $g_{i}$ is the coupling between transmon $i$ and the cavity, $\varepsilon(t)$ is the time-dependent pulse envelope, $\omega_{d}$ is the drive frequency, $a^{(\dagger)}$ are the annihilation (creation) operators of the cavity mode photon, and $c^{(\dagger)}$ are the annihilation (creation) operators of the transmon excitations.

In order to investigate the two-qubit operation of interest, it is instructive to first look at a case in which the transmon is approximated as a two-level system. This will be less computationally expensive in numerical optimization calculations, and will give an indication as to the effectiveness of these techniques in such a system while also revealing the single- and two-qubit interactions during the operation. The Hamiltonian for two two-level systems and a microwave drive coupled to a common cavity mode is given by:

\begin{equation}
\begin{split}
    H= \hspace{1mm}&\omega_{r}a^{\dagger}a+\sum_{j=1,2}\frac{\omega_{a}^{(j)}}{2}\sigma_{z}^{(j)}+\sum_{j=1,2}g_{j}\sigma_{x}^{(j)}(a^{\dagger}+a)\\
    &+\varepsilon(t)(a^{\dagger}e^{-i\omega_{d}t}+ae^{+i\omega_{d}t}),
\end{split}
\end{equation}
\\
where $j$ denotes qubits 1 and 2, $\omega_{a}^{(j)}$ is qubit $j$ transition frequency, $g_{j}$ is the coupling between qubit $j$ and the cavity, and $\sigma_{x/z}^{(j)}$ are the Pauli spin matrices for qubit $j$. Here we have assumed a real drive.

When the drive is far detuned from the cavity, as is the case here, the effect of the drive on the cavity is not important. In order to see the effect of the drive on the qubits, the time-dependent displacement operator $D(\alpha)=\mathrm{exp}(\alpha a^{\dagger}-\alpha^{*}a)$ is be applied \cite{Blais2007}, choosing $\alpha$ to satisfy:

\begin{equation}
    \omega_{r}\alpha +\varepsilon e^{-i\omega_{d}t}-i\dot{\alpha}=0,
\end{equation}
\\
this Hamiltonian becomes:

\begin{equation}
\begin{split}
    H'=&\omega_{r}a^{\dagger}a+\sum_{j=1,2}\frac{\omega_{a}^{(j)}}{2}\sigma_{z}^{(j)}+\sum_{j=1,2}g_{j}\sigma_{x}^{(j)}(a^{\dagger}+a)\\
    &+\sum_{j=1,2}\Omega_{R}^{(j)}\sigma_{x}^{(j)}\mathrm{cos}(\omega_{d}t),
\end{split}
\end{equation}
\\
where $\Omega_{R}^{(j)}=2g_{j}\varepsilon/(\omega_{r}-\omega_{d})$. This shows an important difference to \cite{Chow2011}, where each drive term would only apply to the qubit it was localized to. In this case, the single-microwave control drives both qubits.

Since the qubits are far detuned from the cavity, with $g_{j}\ll|\Delta_{j}|=|\omega_{a}^{(j)}-\omega_{r}|$, the dispersive transformation can be applied:

\begin{equation}
    U=\mathrm{exp}\bigg[\frac{g_{1}}{\Delta_{1}}(a^{\dagger}\sigma_{-_{1}}-a\sigma_{+_{1}})+\frac{g_{2}}{\Delta_{2}}(a^{\dagger}\sigma_{-_{2}}-a\sigma_{+_{2}})\bigg].
\end{equation}
\\
Expanding to second order in the small parameter $g_{j}/\Delta_{j}$ and dropping fast oscillating terms gives the effective Hamiltonian:

\begin{equation}\label{H''}
\begin{split}
    H''=&\omega_{r}a^{\dagger}a+\sum_{j=1,2}\frac{\omega'{}_{a}^{(j)}}{2}\sigma_{z}^{(j)}\\
    &+\frac{g_{1}g_{2}(\Delta_{1}+\Delta_{2})}{2\Delta_{1}\Delta_{2}}(\sigma_{+}^{(1)}\sigma_{-}^{(2)}+\sigma_{-}^{(1)}\sigma_{+}^{(2)})\\
    &+\sum_{j=1,2}\Omega_{R}^{(j)}\sigma_{x}^{(j)}\mathrm{cos}(\omega_{d}t),
\end{split}
\end{equation}
\\
where $\omega'{}_{a}^{(j)}$ is the shifted qubit frequency given by:

\begin{equation}
    \omega'{}_{a}^{(j)}=\omega_{a}^{(j)}+2\frac{g_{j}^{2}}{\Delta_{j}}\bigg(a^{\dagger}a+\frac{1}{2}\bigg).
\end{equation}
\\
This is now the Hamiltonian for two coupled qubits with a single-microwave drive that drives both qubits. 

Following Ref. \cite{Richer2013}, the final step is to perform a Schrieffer-Wolff transformation which transforms the effective Hamiltonian via:

\begin{equation}
    H^{eff}=H^{0}+\frac{1}{2}[H^{2},S^{(1)}]+H_{d}(t)+[H_{d}(t),S^{(1)}],
\end{equation}
\\
where $H^{0}$ is the unperturbed part of the Hamiltonian [given on the first line of Eq. (\ref{H''})], $H^{2}$ is the small perturbation term that contains off-diagonal terms [second line of Eq. (\ref{H''})], and $H_{d}(t)$ is the drive term [third line of Eq. (\ref{H''})]. For this derivation, $S^{(1)}$ is given by:

\begin{equation}
    S^{(1)}=-\frac{J}{\Delta_{12}}(\sigma_{+_{1}}\sigma_{-_{2}}-\sigma_{-_{1}}\sigma_{+_{2}}),
\end{equation}
\\
with $J=g_{1}g_{2}(\Delta_{1}+\Delta_{2})/2\Delta_{1}\Delta_{2}$ and $\Delta_{12}=\omega_{a_{1}}'-\omega_{a_{2}}'$, where $J\ll\Delta_{12}$. The final Hamiltonian is then given by:

\begin{equation}\label{CRHam}
\begin{split}
    H^{eff}=&\omega_{r}a^{\dagger}a+\frac{\tilde{\omega}_{a}^{(1)}}{2}\sigma_{z}^{(1)}+\frac{\tilde{\omega}_{a}^{(2)}}{2}\sigma_{z}^{(2)}\\
    &+\Omega_{R}^{(1)}\bigg(\sigma_{x}^{(1)}+\frac{J}{\Delta_{12}}\sigma_{z}^{(1)}\sigma_{x}^{(2)}\bigg)\mathrm{cos}(\omega_{d}t)\\
    &+\Omega_{R}^{(2)}\bigg(\sigma_{x}^{(2)}-\frac{J}{\Delta_{12}}\sigma_{x}^{(1)}\sigma_{z}^{(2)}\bigg)\mathrm{cos}(\omega_{d}t),
\end{split}
\end{equation}
\\
where $\tilde{\omega}_{a}^{(1)}/2=\omega'{}_{a}^{(1)}+J^{2}/\Delta_{12}$, and $\tilde{\omega}_{a}^{(2)}=\omega'{}_{a}^{(2)}-J^{2}/\Delta_{12}$. This Hamiltonian contains the two-qubit terms $\sigma_{z}^{(1)}\sigma_{x}^{(2)}$ and $\sigma_{x}^{(1)}\sigma_{z}^{(2)}$. To activate one of these terms a drive at the correct frequency must be applied. In order to perform the $\sigma_{x}^{(1)}\sigma_{z}^{(2)}$ operation, for example, a microwave drive at $\omega_{d}=\tilde{\omega}_{a}^{(1)}$ must be applied. This $\sigma_{x}^{(1)}\sigma_{z}^{(2)}$ is two single-qubit rotations away from being a controlled-not (CNOT) gate. 

According to Eq. (\ref{CRHam}), choosing $\omega_{d}=\tilde{\omega}_{a}^{(1)}$ will not only perform the $\sigma_{x}^{(1)}\sigma_{z}^{(2)}$ operation, this will also drive a single-qubit rotation on qubit 1 while also driving some off-resonant rotations of qubit 2 and the $\sigma_{z}^{(1)}\sigma_{x}^{(2)}$ term. However, performing the relevant single- or two-qubit operation is not as simple as just choosing the correct frequency and driving, the pulse shape also plays an important part. For example, choosing to drive at $\omega_{d}=\tilde{\omega}_{a}^{(1)}$ using the pulse shape shown in Fig. \ref{InitPulse16025Gauss} would achieve a fidelity of $\mathcal{F}=0.1461$ for a $\pi/2$ rotation on the $\sigma_{x_{1}}\sigma_{z_{2}}$ operation, and $\mathcal{F}=0.3230$ for a $\pi/2$ rotation on the $\sigma_{x_{1}}$ operation.

In very few cases, solutions for the problem of finding pulse shapes to perform required operations can be found analytically. In most cases, however, such as here, this is not possible, as the solutions must be found numerically. There may also exist many solutions to the problem; this makes finding a solution non-trivial.

\section{\label{sec:OCT}Quantum Optimal Control}

The aim of optimal control theory is to maximize or minimize a function subject to certain constraints and bounds. For quantum information processing the function we wish to maximize is the fidelity of the system unitary evolution after some time $T$, $U(T)$, with respect to a desired target unitary operator, $W\in\mathbb{C}^{n_{s}\times n_{s}}$ where $n_{s}$ is the size of the system. In quantum systems the evolution of the system can be described by:

\begin{equation}\label{OCTeqn}
    i\dot{U}(t)=\bigg(H_{0}+\sum_{j}c_{j}(t)H_{j}\bigg)U(t),
\end{equation}
\\
with $H_{0}$ the drift (i.e. not controlled) Hamiltonian, $j$ the index of the control, $c_{j}(t)$ the control function at time $t\in[0,T]$, and $H_{j}$ the control Hamiltonian that is under the influence of control $j$. The fidelity function used here is:

\begin{equation}
    \mathcal{F}=\bigg\lvert\frac{1}{n_{s}}\mathrm{tr}[W^{\dagger}\hat{O}U(T)\hat{O}]\bigg\rvert^{2},
\end{equation}
\\
where $\hat{O}$ is the projector onto the subspace of interest. With this definition we have $\mathcal{F}\in [0,1]$, when $\mathcal{F}=1$ there is no measurable distinction between $W$ and $U(T)$, and no leakage out of the relevant subspace.

In general, maximizing the fidelity function is performed numerically and can be very computationally expensive, particularly when using smooth analytic control functions, due to the large size of the control space (i.e. the dimension of the control space). For the purposes of optimization it can be more efficient to approximate the controls as a series of piecewise constant amplitudes over $N$ uniform time intervals $\tau=T/N$, giving the Hamiltonian for time $t_{k}=k\tau$ in Eq. (\eqref{OCTeqn}) as:

\begin{equation}\label{OCHam}
    H(t_{k})=H_{0}+\sum_{j}c_{j}(t_{k})H_{j},\hspace{2.5mm} t_{k}=k\tau.
\end{equation}
\\
In this case, and with $H_{j}$ time-independent, $U(T)$ is now given by $U(T)=\prod_{k=1}^{N}U_{k}$ with $U_{k}=\mathrm{exp}\big(-iH(t_{k})\tau\big)$.

The most successful methods for finding optimal controls have been local optimizers based on gradient ascent \cite{Navin2005, Brif2010, Gross2007}. For these algorithms an initial control pulse is chosen, the fidelity and the gradient are calculated and the optimizer is run which uses the gradient at each iteration of the optimization to determine in which direction in the landscape to travel. These algorithms have been shown to converge on a solution in fewer iterations than other algorithms and are computationally less expensive, however they do not guarantee a global maximum as this depends on where you start in the control landscape.

When applying optimal control to quantum information problems the pulses that are developed must be robust to any errors in the system, such as errors in the measured system parameters. In order to perform numerical optimal control simulations the system parameters must be entered into the simulator, but this requires the parameters to be known with absolute precision which will not be the case in experiment. The estimated parameters will come with some error range, which must be included in the numerical calculation. As discussed in \cite{Kosut2013}, a method for doing this is to sample points $\delta$ from the parameter range $\Delta$, calculate the fidelity for each of these points and then to maximize the worst-case fidelity, $\mathcal{F}_{wc}$: 

\begin{equation}
\begin{split}
      &\mathrm{maximize}\hspace{5mm}\mathcal{F}_{wc}=\mathrm{min}_{\delta}\mathcal{F}(\theta,\delta)\\
      &\mathrm{subject}\hspace{1mm}\mathrm{to}\hspace{5mm}\theta\in\Theta,\hspace{1mm}\delta\in\Delta,
\end{split}
\end{equation}
\\
where $\theta$ are the controls and $\Theta$ is the set of allowed controls subject to any constraints imposed. Maximizing the minimum fidelity of the range, rather than the average fidelity of the range, places a more strict requirement on the optimizer and ensures that the control found by the optimizer does not include points with very low fidelity that are balanced by high fidelity terms.

\subsection*{\label{SCP}Sequential convex programming}

The method proposed for solving the optimization problem stated above is sequential convex programming \cite{Schittkowski1995a, Kosut2013}. This is a gradient based local optimizer and is therefore efficient at locating local optimal solutions, provided the initial guess is a good one. The algorithm is initialized with a control $\theta\in\Theta$, which is typically a convex set, or well approximated by one. Points $\delta_{i}$ are then sampled from the error range $\Delta$, and a convex trust region $\tilde{\Theta}_{trust}$ is initialized. The trust region is chosen so that the linearized fidelity $\mathcal{F}(\theta,\delta_{i})+\tilde{\theta}^{\intercal}\nabla_{\theta}\mathcal{F}(\theta,\delta_{i})$, where $\tilde{\theta}\in\tilde{\Theta}_{trust}$, used in the optimization step retains sufficient accuracy. 

With the initial points set, the fidelities $\mathcal{F}(\theta,\delta_{i})$ and gradients $\nabla_{\theta}\mathcal{F}(\theta,\delta_{i})$ are calculated for each points $\delta_{i}$ selected from the error range. The linearised fidelity is then used to solve for the increment $\tilde{\theta}$ in the convex optimisation:

\begin{equation}
    \begin{split}
        &\mathrm{maximise}\hspace{2.5mm}\mathrm{min}_{i}[\mathcal{F}(\theta,\delta_{i})+\tilde{\theta}^{\intercal}\nabla_{\theta}\mathcal{F}(\theta,\delta_{i})]\\
        &\mathrm{subject}\hspace{1mm}\mathrm{to}\hspace{2.5mm}\theta+\tilde{\theta}\in\Theta,\hspace{1mm}\tilde{\theta}\in\Tilde{\Theta}_{trust}.
    \end{split}
\end{equation}
\\
If $\mathrm{min}_{i}\mathcal{F}(\theta+\tilde{\theta},\delta_{i})>\mathrm{min}_{i}\mathcal{F}(\theta,\delta_{i})$ then replace $\theta$ by $\theta+\tilde{\theta}$, increase the trust-region $\tilde{\Theta}_{trust}$ and repeat the process of calculating fidelities, gradients and solving for the increment $\tilde{\theta}$. If, however, the $\mathrm{min}_{i}\mathcal{F}(\theta+\tilde{\theta},\delta_{i})<\mathrm{min}_{i}\mathcal{F}(\theta,\delta_{i})$ then decrease the trust region $\tilde{\Theta}_{trust}$ and repeat the optimisation step with the same $\theta$. This process is repeated until some stopping criteria are satisfied, such as the number of iterations reaching the maximum number imposed or the gradient is below some threshold such that it is flat and has thus found a local maximum.

\section{Cross Resonance Using SCP}\label{sec:Results}

To perform quantum optimal control the effective Hamiltonian in eqn. \eqref{CRHam} must be cast in the form of eqn. \eqref{OCHam}. This investigation focuses on having all the system parameters fixed and the control variable as the pulse shape $\varepsilon(t)$. To begin the investigation, we first look at the Hamiltonian in the two-level approximation. Moving to a frame rotating at the drive frequency, the drift Hamiltonian is given by:

\begin{equation}
    H_{0}=\Delta_{r}a^{\dagger}a+\sum_{j=1,2}\frac{\tilde{\Delta}_{a}^{(j)}}{2}\sigma_{z_{j}},
\end{equation}
\\
where $\Delta_{r}=\omega_{r}-\omega_{a}$ and $\tilde{\Delta}_{a_{j}}=\tilde{\omega}_{a}^{(j)}-\omega_{d}$. The control Hamiltonian is given by:

\begin{equation}
\begin{split}
    H_{j}=&\frac{2g_{1}}{\omega_{r}-\omega_{d}}\bigg(\sigma_{x_{1}}+\frac{J}{\Delta_{12}}\sigma_{z_{1}}\sigma_{x_{2}}\bigg)\\
    &+\frac{2g_{2}}{\omega_{r}-\omega_{d}}\bigg(\sigma_{x_{2}}-\frac{J}{\Delta_{12}}\sigma_{x_{1}}\sigma_{z_{2}}\bigg),
\end{split}
\end{equation}
\\
and thus $c(t)=\varepsilon(t)$. In these simulations the system parameters are set as $\omega_{r}/2\pi=6.44$ GHz, $\omega_{a}^{(1)}/2\pi=4.50$ GHz, $\omega_{a}^{(2)}/2\pi=4.85$ GHz, $g_{1}/2\pi=g_{2}/2\pi=133$ MHz; these are parameters that have been used in a previous cQED experiment \cite{Leek2009}. The desired unitary is:

\begin{equation}
    U_{des}=\mathrm{exp}\bigg(-i\frac{\pi}{4}\sigma_{x_{1}}\sigma_{z_{2}}\bigg),
\end{equation}
\\
which, when applied to the state $\ket{+y}\ket{+y}$ [where $\ket{+y}=(\ket{0}+i\ket{1})/\sqrt{2}$], gives the maximally entangled Bell state $\ket{\Phi^{-}}=(\ket{00}-\ket{11})/\sqrt{2}$. In order to perform this, we choose $\omega_{d}=\tilde{\omega}_{a_{1}}$.

To ensure realistic pulses are produced, constraints on the maximum and minimum amplitudes have been imposed. By doing this the control space becomes fragmented, as there will be areas that are off limits to the optimizer. This means that the control space may contain many local maxima that the optimizer may become stuck in, as the optimization problem is very sensitive to initial conditions.

\begin{figure}[h!]
    \centering
    \includegraphics[width=0.4\textwidth]{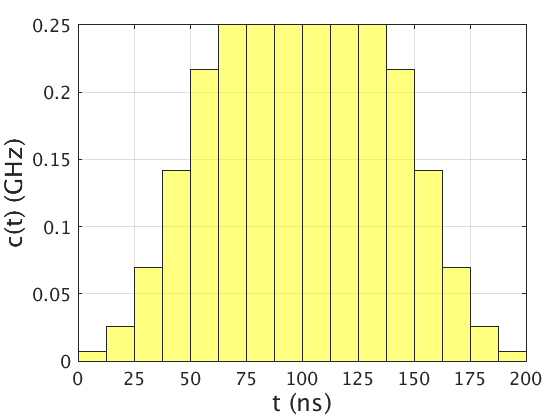}
    \caption{Example of a pulse shape where the pulse has been broken up into 16 piecewise constant parts over an interval of $12.5$ ns each. $c(t)=\varepsilon(t)/2\pi$ gives the pulse amplitude in ${\rm GHz}$ and is over an interval of $200$ ns total time. Each separate part can be varied to optimize over some problem. This one takes the form of a Gaussian turn on, with a flat top and a Gaussian turn off.}
    \label{InitPulse16025Gauss}
\end{figure}

In \cite{Chow2011}, it was stated the pulse used to perform the cross-resonance gate was a slow Gaussian turn on with a flat top. Therefore, as an initial pulse guess for the SCP algorithm we have chosen to use a pulse that has the form of a Gaussian turn on, a flat top, and then a Gaussian turn off, as shown in Fig. \ref{InitPulse16025Gauss}. 

\begin{figure}[h!]
    \centering
    \includegraphics[width=0.4\textwidth]{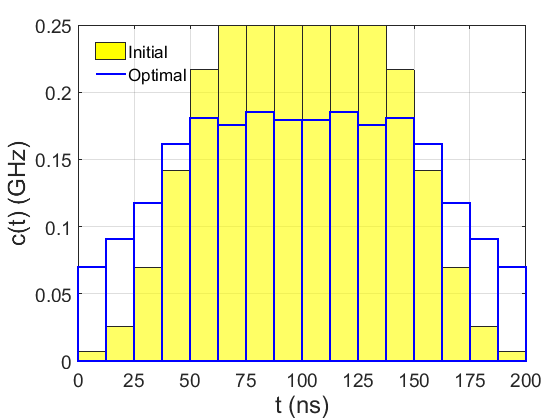}
    \caption{The initial pulse sequence for the SCP algorithm (coloured area), with $\mathcal{F}=0.1461$, and the optimal pulse sequence (outlined area) showing the variation from initial to final, when the pulse is broken into 16 piecewise constant parts. As is shown, the solution tends to stay close to the initial solution if a good initial guess is chosen. Here, $c(t)=\varepsilon(t)/2\pi$ is given in $\mathrm{GHz}$ while $\mathrm{t}$ is given in $\mathrm{ns}$. This optimal pulse sequence generates the desired unitary with $\mathcal{F}=0.9945$.}
    \label{OptPulse02516}
\end{figure}

\begin{figure}[t]
    \centering
\includegraphics[width=0.4\textwidth]{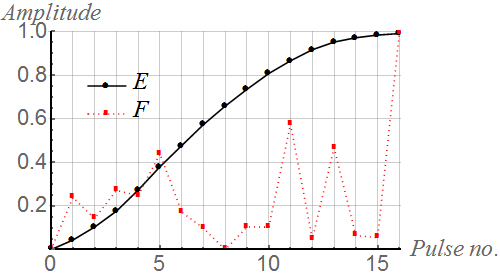}
    \caption{ 16 pulses are applied over an interval of $\tau=12.5$ ns each. The entanglement of a pure state is given by $2|ad-bc|$ for an arbitrary two-qubit state $\ket{\psi} = a\ket{00}+b\ket{01}+c\ket{10}+d\ket{11}$. The fidelity curve is given by $F=|\bra{\Phi^{-}} \tilde{U}^{m} \ket{+y} \ket{+y}|^2$ where $\tilde{U}^{m}\ket{+y}\ket{+y}=\Pi^{m}_{k=1}exp\big(-iH(t_{k})\tau\big)\ket{+y}\ket{+y} $. }
    \label{JaewooGraph}
\end{figure}

From these initial conditions, with $\mathcal{F}=0.1461$, a pulse sequence can be generated that is able to perform the desired unitary with a fidelity of $\mathcal{F}=0.9945$. Figure \ref{OptPulse02516} shows the changes that have been made from the original pulse sequence, shaded area, to the optimal pulse shape, outlined area. Due to a good choice of initial guess the optimiser hugs this shape and rapidly finds a solution with high fidelity, even with the constraint on maximum amplitude and few pulse pixels to optimize over. Figure \ref{JaewooGraph} shows, with respect to the state:

\begin{equation}
\tilde{U}^{m}\ket{+y}\ket{+y}=\Pi^{m}_{k=1}exp\big(-iH(t_{k})\tau\big)\ket{+y}\ket{+y},
\end{equation}
\\
how the entanglement (given by $2|ad-bc|$ for an arbitrary two-qubit state $\ket{\psi}=a\ket{00}+b\ket{01}+c\ket{10}+d\ket{11}$) and the fidelity (given by $\mathcal{F}=|\bra{\Phi_{-}}\tilde{U}^{m}\ket{+y}\ket{+y}|^{2}$) change with each successive $c(t_{k})$. We see that although the fidelity fluctuates, the entanglement monotonically increases with each pulse. For example, at $k=8$, the entanglement is higher than 0.6 but the fidelity is nearly 0 because the outcome state is given by

\begin{equation}
\begin{split}
&\tilde{U}^{8} \ket{+y} \ket{+y}  \approx \alpha \ket{\Phi^{+}} + \beta \ket{\Psi^{-}}, \\
&F \approx |\bra{\Phi^{-}} \left(\alpha \ket{\Phi^{+}} + \beta \ket{\Psi^{-}} \right) |^2 =0, 
\label{8th}
\end{split}
\end{equation}
\\
where $\ket{\Phi^{+}}=(\ket{00}+\ket{11})/\sqrt{2}$, $\ket{\Psi^{-}}=(\ket{01}-\ket{10})/\sqrt{2}$, and $\alpha, \beta\in\mathbb{C}$. This suggests that the pulse shape continually performs the desired two-qubit operations, but has optimised to produce single-qubit rotations at each step such that at the end the qubits will be rotated into the correct basis and that the function performed at the end is effectively just the desired two-qubit operation.

\begin{figure}[h!]
    \centering
    \includegraphics[width=0.4\textwidth]{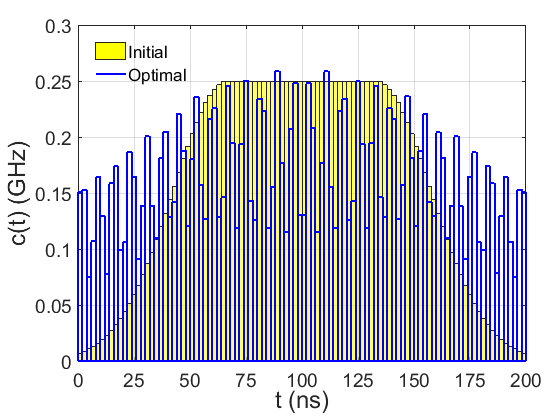}
    \caption{The initial (coloured area) and optimal (outlined area) pulses when the control is split up into $100$ piecewise constant parts, each over an interval of $2~{\rm ns}$ which is close to the limit of microwave generators. In this case the final pulse, while showing a similar overall structure to the initial pulse, has large amplitude changes between each piece. This could off resonantly drive unwanted higher frequency terms and cause problems in actual implementation. $c(t)=\varepsilon(t)/2\pi$ is in $\mathrm{GHz}$ and $\mathrm{t}$ is in $\mathrm{ns}$. The fidelity that this pulse generates is $\mathcal{F}=0.9940$.}
    \label{Nt100overlay}
\end{figure}

For the purposes of our optimization, the assumption of piecewise constant controls has been made. This is to make the optimization less computationally expensive as there will be fewer parameters to optimize; the control space dimension will be given by the number of piecewise constant parts in the pulse. It has been suggested that piecewise constant controls are not as powerful as analytic controls, and that analytic controls should always be used where possible \cite{Machnes2015}. However, the assumption was made that the piecewise constant controls were used as an approximation of a continuous function and thus would need many time slices to accurately portray such a function. This then made the optimization very computationally expensive. Here instead we choose our control amplitudes, not as an approximation of a continuous function, but as a stand alone pulse shape. State-of-the-art arbitrary waveform generators (AWGs) can generate pulses that change amplitude at least on the order of $1$ ns. We found that the optimal choice of pulse with fewer piecewise constant parts was a pulse over $200$ ns that has 16 individual constant pulses that can be optimized, so that each constant pulse is over an interval of $12.5$ ns. This is well within the capability of many AWGs, and allows us to carry out a resource-efficient optimization. Use of pulses with shorter time steps, to approximate a continuous function, does not necessarily improve the fidelity. In Fig. \ref{Nt100overlay}, a pulse sequence generated is shown where each pulse length is 2 ns, getting closer to the limit of the AWGs, which implements the desired unitary with $\mathcal{F}=0.9940$. This optimisation performed the calculations around 12 times more slowly than with fewer piecewise constant parts (3 hours for the larger control space compared with 15 minutes for the smaller one). 

\subsection*{Optimal pulses with filtering effect}

In general the pulse shown in figure \ref{OptPulse02516} will not be the pulse that reaches the cavity and qubits. The microwave pulse will be (mostly low pass) filtered by control hardware; we can model this by discretizing the control further and using a linear transfer function to approximate the filtering effect \cite{Motzoi2011}. In this case, each new piecewise constant pixels $s(t_{l})$ are given by:

\begin{equation}
    s(t_{l})=\sum_{k=0}^{N-1}T_{l,k}c(t_{k}),
\end{equation}
\\
where $T_{l,k}$ is the transfer function as given in \cite{Motzoi2011} for a Gaussian filter.

\begin{figure}[h!]
    \centering
    \includegraphics[width=0.4\textwidth]{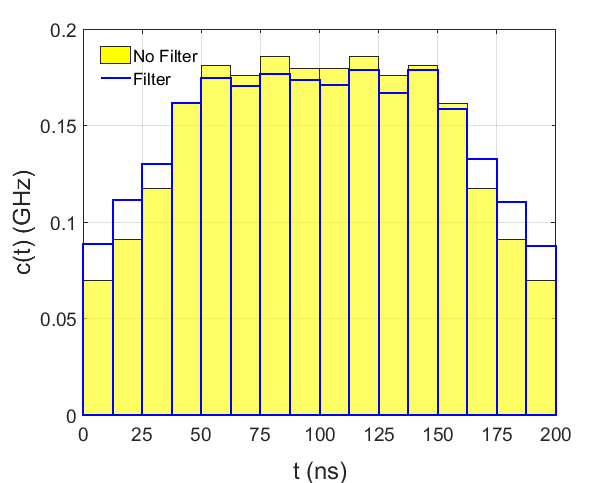}
    \caption{The optimal pulse when there is no filtering (coloured area) gives a fidelity of $\mathcal{F}=0.8303$ when the simulation takes into account filtering. Upon reoptimizing, a new pulse is generated (shaded area) which generates the desired unitary with $\mathcal{F}=0.9947$.}
    \label{pulsetau005noerror}
\end{figure}

This is applied to the pulse in Fig. \ref{OptPulse02516}, the fidelity calculated drops to $\mathcal{F}=0.8303$, but upon re-optimizing with this pulse as the initial guess a fidelity of $\mathcal{F}=0.9947$ is achieved. Figure \ref{pulsetau005noerror} shows the optimal pulse before re-optimizing (colored area) and the new pulse generated after the filter effect is taken into account (outlined area). Thus, pulses that are robust to any filtering effect may be produced by the algorithm, which is ideal for experimental implementation.

\subsection*{Robust pulses for errors in system parameters}

\begin{figure}[h!]
    \centering
    \includegraphics[width=0.4\textwidth]{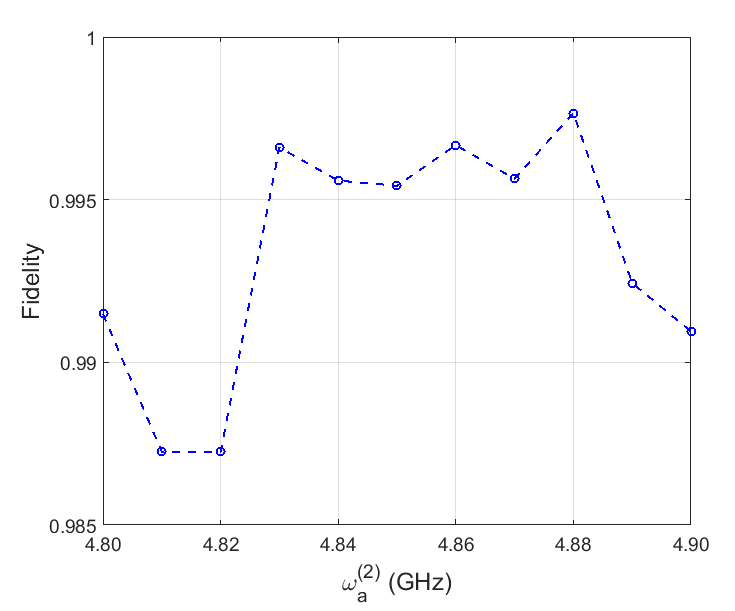}
    \caption{Fidelity against $\omega_{a}^{(2)}/2\pi$ when there is no filter in the simulation. It can be seen that the fidelity for each point $i$ $\mathcal{F}_{i}>0.986$, and that in the range $4.83$ GHz $<\omega_{a}^{(2)}/2\pi<4.88$ GHz the fidelities are all $\mathcal{F}_{i}>0.995$. Therefore a robust pulse has been generated for the range of qubit 2 values: $\omega_{a}^{(2)}/2\pi=4.85\pm0.5$ GHz.}
    \label{fidelitywa2nofilter}
\end{figure}

\begin{figure}[h!]
    \centering
    \includegraphics[width=0.4\textwidth]{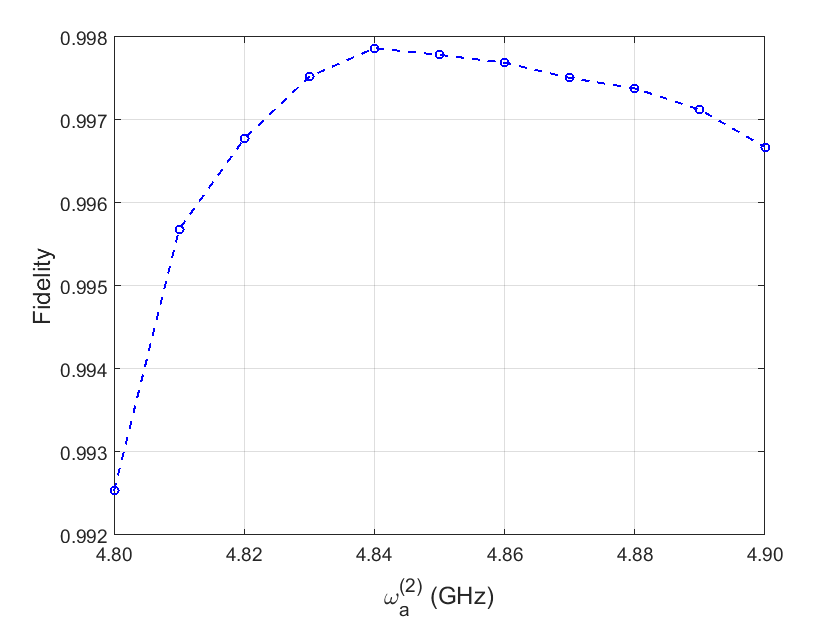}
    \caption{Fidelity against $\omega_{a}^{(2)}/2\pi$ when the filter effect has been taken into account. In this case a robust pulse has been generated that gives $\mathcal{F}_{i}>0.992$ for all points in the range $\omega_{a}^{(2)}/2\pi=4.85\pm0.5~{\rm GHz}$.}
    \label{fidelitywa2filter}
\end{figure}

As mentioned in Sec. \ref{sec:OCT}, the values of the system parameters are going to have some error associated with them, as any measured system parameter is given with some error range. A sample from the error ranges must be taken and the SCP algorithm can be performed while optimizing for the minimum fidelity in the range. This is similar to the approach taken in \cite{Dong2014a}, except that the more stringent condition of maximizing the minimum fidelity is made, as in \cite{Kosut2013}, rather than optimizing for the average fidelity. This method is first performed for an error of $\pm1\%$ in the transition frequency of qubit 2, $\omega_{a}^{(2)}/2\pi=4.85\pm0.05$ GHz. 11 points are sampled from the error range, which has been shown to be adequate to cover the range \cite{Dong2014a}. Figure \ref{fidelitywa2nofilter} shows the range of fidelities for each point sampled from the error range of $\omega_{a}^{(2)}$. This shows that even in the presence of error in one parameter the SCP algorithm has been able to find a solution that can produce the desired unitary with $\mathcal{F}>0.986$ for the whole range. When filtering is included a solution is found which gives the range of fidelities in figure \ref{fidelitywa2filter}.

\begin{figure}[h!]
    \centering
    \includegraphics[width=0.4\textwidth]{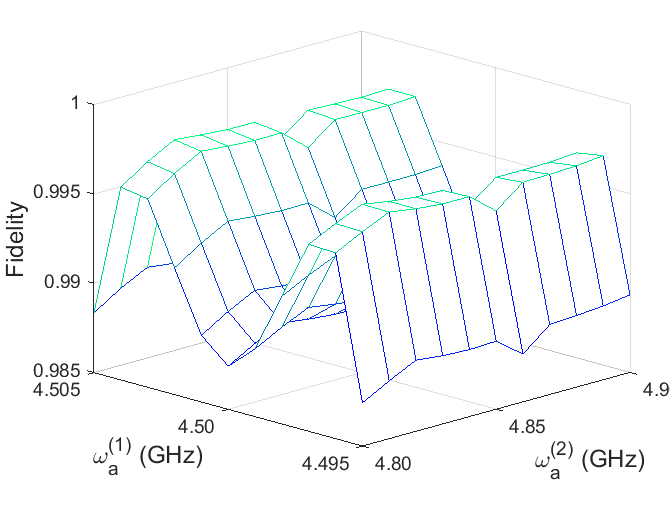}
    \caption{Fidelity against $\omega_{a}^{(1)}/2\pi$ and $\omega_{a}^{(2)}/2\pi$ when filtering is turned on in the simulation. Here we have $\mathcal{F}_{i}>0.9875$ for all points $i$ in the ranges of $\omega_{a}^{(2)}/2\pi=4.85\pm0.05$ GHz, $\omega_{a}^{(1)}/2\pi=4.50\pm0.005$ GHz.}
    \label{wa1wa2upclose2}
\end{figure}

Since there are two qubits, potential errors in both of the qubits parameters, $\omega_{a}^{(1)}$ and $\omega_{a}^{(2)}$, must be accounted for. This proves to be a greater challenge since the drive is on resonance with the dressed qubit-1 transition frequency and thus will cause the drive to become slightly off resonant if there are errors. Reducing the error range for $\omega_{a}^{(1)}$ to $\pm~0.1\%$ achieves a solution with the range of fidelities shown in Fig. \ref{wa1wa2upclose2}, where all points have fidelity $\mathcal{F}>0.9875$. Using the robustness method can be effective in designing a pulse to be robust to errors in some range, but that it becomes increasingly difficult as more error parameters are introduced. This can perhaps be solved by choosing different start points for the optimiser, but thus far the flat-top-like Gaussian has proven to be the optimal start point for the Hamiltonian and desired unitary of interest. Future work will look at a scheme to over come this.

\subsection*{Multi-level Transmon Model}

We now proceed to look at the case in which a full model of the transmons is incorporated. This is more computationally expensive to calculate, but is important for capturing leakage out of the two-level logical basis. In the deep transmon limit, with $E_{J}/E_{C}=100$, the anharmonicities for the transmons are $\delta_{1}/2\pi=-160$ MHz for transmon 1, and $\delta_{2}/2\pi=-170$ MHz for transmon 2. In this limit, the transmons can be approximated as Duffing oscillators, the Hamiltonian for two Duffing oscillators coupled to a common cavity mode with a single drive is given by:

\begin{equation}\label{Duffing}
\begin{split}
    H=&\omega_{r}a^{\dagger}a+\sum_{j=1,2}\bigg(\omega_{a}^{(j)}c_{j}^{\dagger}c_{j}+\frac{\delta_{j}}{2}c_{j}^{\dagger}c_{j}(c_{j}^{\dagger}c_{j}-1)\bigg)\\
    &+\sum_{j=1,2}g_{j}(a^{\dagger}c_{j}+ac_{j}^{\dagger})\\
    &+(\varepsilon(t)a^{\dagger}e^{-i\omega_{d}t}+\varepsilon^{*}(t)ae^{+i\omega_{d}t}).
\end{split}
\end{equation}
\\
The first two lines of this equation can be diagonalised to find the Hamiltonian for the cavity + two transmons in the dressed basis, where the frequencies of each component will now include dependencies on all the other parts. For the purposes of optimal control this will now form the drift Hamiltonian. The drive term can then be transformed to form the new operators in the dressed basis, and can be used as the new drive term $H_{c}$ for the simulations.

For the multi-level simulations the drive term has been cast into two parts to include complex control as it has been shown that using both quadratures can be useful in suppressing leakage \cite{Motzoi2009}. The drive term in this case becomes:

\begin{equation}
\begin{split}
H_{d} = &\varepsilon_{x}(t)(a^{\dagger}e^{-i\omega_{d}t}+ae^{+i\omega_{d}t})\\
&+i\varepsilon_{y}(t)(a^{\dagger}e^{-i\omega_{d}t}-ae^{+i\omega_{d}t}).
\end{split}
\end{equation}

\begin{figure}[h!]
    \centering
    \includegraphics[width=0.4\textwidth]{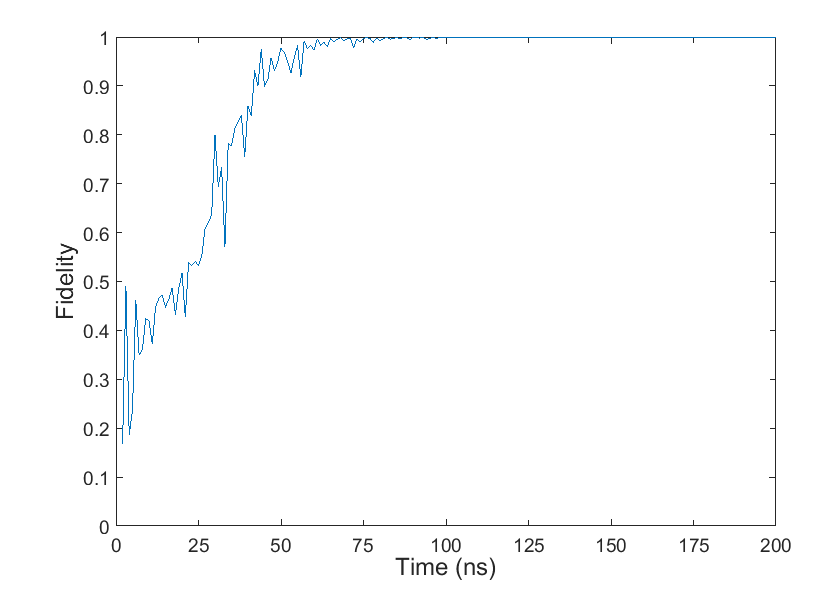}
    \caption{Fidelity of optimised pulses using a multi-level system Hamiltonian comprised of two Duffing oscillators each coupled to a common cavity mode with a single cavity drive. The initial pulse had the form of a flat-top Gaussian with each piecewise constant part being ~2ns long, for different total times ranging from 2 to 200 ns. For $T<100$ ns the optimised pulses perform poorly, however for all times $T>100$ ns the fidelity converges to $\mathcal{F}>0.999$ and even to $\mathcal{F}>0.9999$ for certain times in this range.}
    \label{FidTimeNt2Flattop}
\end{figure}

In the new transmon limit it is that the control must be discretized more in order to reach a good fidelity. In this case each piecewise control amplitude is now $2$ ns long, well within the capabilities of current AWGs. In the case where there are no errors in the system there are many points in time that perform well with fidelities $\mathcal{F}>0.9999$, due to this extra discretization of the control. Figure \ref{FidTimeNt2Flattop} shows the fidelity of the optimised pulses, where each initial pulse has taken the form of a flat-top Gaussian with dt=2 ns for each piecewise amplitude, against time. The fidelities converge to $\mathcal{F}>0.999$ for all times $T>100\mathrm{ns}$.

\begin{figure}[!htbp]
    \centering
    \includegraphics[width=0.4\textwidth]{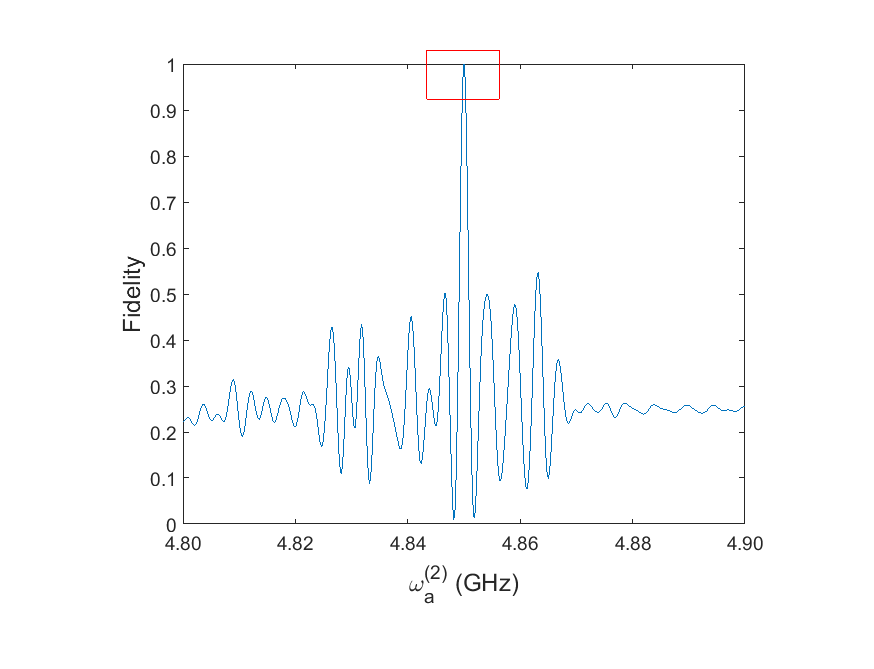}
    \caption{Variation in the fidelity, $\mathcal{F}$, with changing $\omega_{a}^{(2)}$ for a pulse given without taking into account an error range in this parameter during the optimisation. The area of interest is highlighted by the red rectangle: at the ideal parameter, with $\omega_{a}^{(2)}/2\pi=4.85$ GHz, $\mathcal{F}>0.9999$, but the fidelity rapidly decreases as the value moves away from the optimal.}
    \label{wa1wa2F}
\end{figure}

We now investigate the robustness of one of these pulses to parameters variation. In figure \ref{wa1wa2F} we plot the pulse fidelity $\mathcal{F}$ against $\omega_{a}^{(2)}$. While the pulse performs well for the specific chosen parameters, it can clearly be seen that the fidelity is highly dependent on the frequency $\omega_{a}^{(2)}$, hence not robust to variation in this parameter.

\begin{figure}[!htbp]
    \centering
    \includegraphics[width=0.4\textwidth]{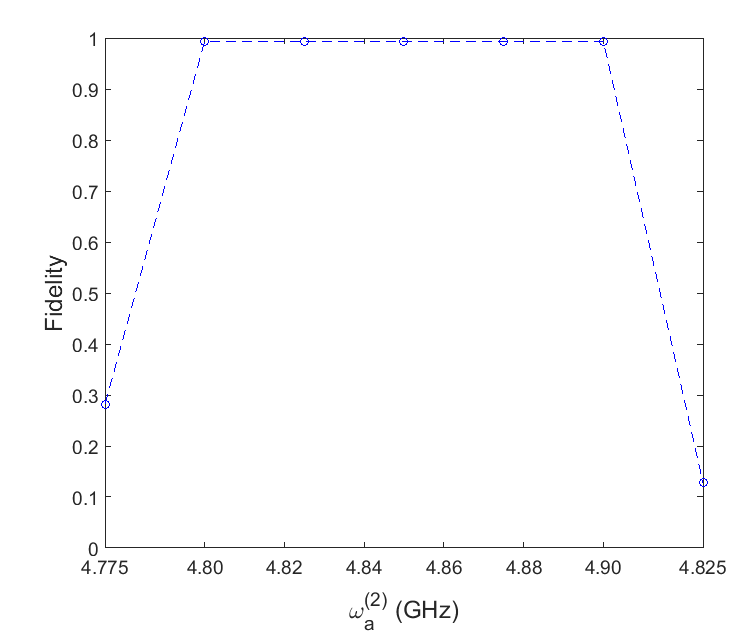}
    \caption{Fidelity against $\omega_{a}^{(2)}/2\pi$ for an optimized pulse using the robust methods on a multi-level system comprised of two Duffing oscillators each coupled to a common cavity mode with a single cavity drive. Here $\mathcal{F}_{i} = 0.9937$ for all points $i$ in the range $\omega_{a}^{(2)}/2\pi=4.85\pm0.05$ GHz, which is the range that has been optimized for. It can be seen that outside of this range the fidelity falls off rapidly.}
    \label{wa2vFT200}
\end{figure}

As in the previous section, the robust methods are used to find a pulse robust to errors in $\omega_{a}^{(2)}$ first. In this case, the algorithm finds a solution that gives a results of $F\approx 0.9937$ for all values in the range $\omega_{a}^{(2)}/2\pi\in [4.80,4.90]$ GHz, shown in Fig. \ref{wa2vFT200}, for a time of $199$ ns. Performing again the robustness method with variations in the two-qubit parameters, a solution is found that is able to perform the desired entangling gate with a fidelity of $\mathcal{F}_{i}\approx 0.9639$ for all parameters in the range $\omega_{a}^{(2)}/2\pi\in [4.80,4.90]$ GHz and $\omega_{a}^{(1)}/2\pi\in [4.495,4.505]$ GHz, for a time of $199$ ns.

In the multilevel case, the same feature appears when attempting to a find a solution that is robust to fluctuations as in the two qubit parameters, i.e., that for just errors in $\omega_{a}^{(2)}$ the algorithm is able to find a solution with $\mathcal{F}>0.99$ for an error of $\pm 1 \%$, but that if we wish to include $\omega_{a}^{(1)}$ it is more difficult to account for this. Nonetheless, with an error of $\pm 0.1 \%$ in the parameter $\omega_{a}^{(1)}$ and $\pm 1\%$ in $\omega_{a}^{(2)}$ a pulse is found that achieves $F>0.96$.

One of the causes of the discrepancy between the fidelities of the two-level case and the multi-level case is the anharmonicity of the transmons we have simulated. Currently, we are operating deep in the transmon regime with $E_{j}/E_{c}=100$ and so one of the limiting factors is down to leakage out of the computational subspace as fluctuations in the qubits $\omega_{01}$ transition bring them even closer to the $\omega_{12}$ transition.

\begin{figure}[!htbp]
    \centering
    \includegraphics[width=0.4\textwidth]{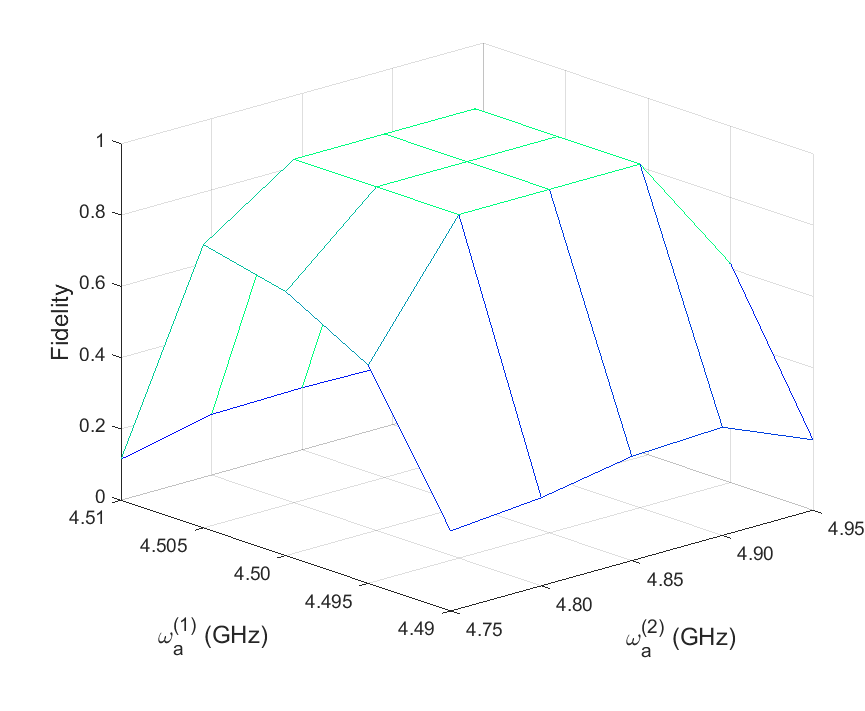}
    \caption{Fidelity against $\omega_{a}^{(2)}/2\pi$, $\omega_{a}^{(1)}/2\pi$ for an optimized pulse using the robust methods on a multi-level system comprised of two Duffing oscillators each coupled to a common cavity mode with a single cavity drive. Here, $\mathcal{F}_{i} = 0.9639$ for all points $i$ in the range $\omega_{a}^{(2)}/2\pi=4.85\pm0.05$ GHz, $\omega_{a}^{(1)}/2\pi=4.50\pm0.005$ GHz, which the pulse has been optimized for. Outside of this range, the fidelity falls off rapidly.}
    \label{wa1wa2FRobust2}
\end{figure}

For this paper, we have chosen not to include errors in the coupling strengths between the transmons and the cavity $g_{j}$. Preliminary tests with errors in coupling strengths have shown minimal effect on the output given without including these errors up to $10 \%$, while this range is feasible in experiments this would merely add to the parameter range selection in the simulations. Errors specifically in the cavity frequency have also not been included since the drive and qubits resonances are far off resonant from the cavity. 

This paper focuses on achieving high-fidelity controls on time scales much shorter than decoherence times and therefore we do not include decoherence in our simulations. Obviously, it is important to correct control errors in this regime if they represent the largest source of infidelity. For example, a state-of-the-art circuit with two transmons with $50$-$\mu s$ coherence times acted on with an entangling unitary operation that requires $200$ ns will see a probability of corruption of the operation due to decoherence estimated at $0.4\%$. Since typical operations errors due unoptimized controls will be larger than this we can focus on optimizing without including decoherence. Usually this is the only relevant regime in which we will gain by optimizing. However, since some forms of decoherence can be tackled actively with dynamical decoupling schemes \cite{Viola1998, Viola1999}, it would be interesting to consider optimizing for Hamiltonian control errors and external decoherence together in the future.

\section{Conclusion}

In conclusion, we have shown that robust quantum control can produce pulse shapes that achieve a desired unitary with high fidelity for a realistic quantum system. In particular, we have shown that in a system where a single-microwave drive coupled to a cavity containing two transmon qubits is chosen on resonance with a qubit, modifying the shape of the driving microwave pulse can produce a desired two-qubit interaction while mitigating the unwanted rotation of the qubit that is also on resonance with the drive. This can be done even in the presence of filtering on the control and errors in the system parameters with a modest amount of resources, and can be achieved even when realistic constraints are placed on the pulses. 

We have seen that including constraints on the pulse opens up more areas of local maxima in the control space, and in this case we found that there were many. These ``traps" may be what is limiting the range of robustness in the two-qubit frequencies. Future work will look at how to make the error range for $\omega_{a}^{(1)}$ larger, potentially by combining these methods with a non-local optimizer in order to circumvent local traps.

We have shown that it is still possible to achieve high fidelity control with reduced circuit complexity, by increasing the complexity of the control. This shows a trade-off between the circuit complexity and pulse complexity, and that as quantum computers grow we are likely to require more complex pulse shapes if we wish to keep the circuit complexity down. Future work will study the limitations of current algorithms to solving these problems.

\section*{Acknowledgements}
E.G. acknowledges financial support from the EPSRC grants (Grants No. EP/L026082/1 and No. EP/L02263X/1). This work was partially supported by the KIST Institutional Program (Project No. 2E26680-16-P025). The data underlying this work are available without restriction. Details of the data and how to request access are available from the University of Surrey publications repository doi:10.15126/surreydata.00813890

\bibliographystyle{apsrev4-1}
\bibliography{bibliography}

\end{document}